\documentclass[referee]{aa}
\usepackage{natbib} \usepackage{graphicx,color}
\bibpunct{(}{)}{;}{a}{}{,}
\begin{document}

\title{Comparison of transient horizontal magnetic fields\\
in a plage region and in the quiet Sun}
\author{R. Ishikawa\inst{1,2} \and S. Tsuneta\inst{2}}
%\email{ryoko.ishikawa@nao.ac.jp}

\institute{Department of Astronomy, University of Tokyo, Hongo, Bunkyo-ku, Tokyo 113-0033, Japan
 \and National Astronomical Observatory of Japan, 2-21-1 Osawa, Mitaka, Tokyo 181-8588, Japan}
\date{Received/Accepted 25 October 2008}

\abstract
{}
%Aim
{Properties of \textit{transient horizontal magnetic fields} (THMFs) in both plage and quiet Sun regions are obtained and compared.}
%Method
{Spectro-polarimetric observations with the Solar Optical Telescope (SOT) on the \emph{Hinode} satellite were carried out with a cadence of about 30 seconds for both plage and quiet regions located near disk center. We select THMFs that have net linear polarization ($LP$) higher than 0.22\%, and an area larger than or equal to 3 pixels, and compare their occurrence rates and distribution of magnetic field azimuth. We obtain probability density functions (PDFs) of magnetic field strength and inclination for both regions.}
%Results
{The occurrence rate in the plage region is the same as for the quiet Sun.
The vertical magnetic flux in the plage region is $\sim$8 times larger than in the quiet Sun.
There is essentially no preferred orientation for the THMFs in either region.
However, THMFs in the plage region with higher $LP$ have a preferred direction consistent with that of the plage-region's large-scale vertical field pattern.
PDFs show that there is no difference in the distribution of field strength of horizontal fields between the quiet Sun and the plage regions when we avoid the persistent large vertical flux concentrations for the plage region.
}
%Conclusion
{The similarity of the PDFs and of the occurrence rates in plage and quiet regions suggests that a local dynamo process due to the granular motion may generate THMFs all over the sun.
The preferred orientation for higher $LP$ in the plage indicates that the THMFs are somewhat influenced by the larger-scale magnetic field pattern of the plage.}
\keywords{Sun: photosphere --- Sun: magnetic field}
\titlerunning{Comparison of THMFs in quiet and plage regions}
\maketitle

\section{Introduction}
Horizontal fields with high time variation occurring all over the Sun were
observed with the SOLIS and GONG instruments \citep{Harvey2007}.
High resolution spectroscopic observations with the SOT spectropolarimeter (SP) \citep{Tsuneta2008SoPh,Suematsu2008SoPh,Ichimoto2008SoPh,Shimizu2008SoPh} aboard \emph{Hinode} \citep{Kosugi2007}
have confirmed this finding and extended the studies considerably \citep{Lites2007,Orozco2007}.
Recently \citet{Ishikawa2008} have reported the presence of THMFs in a plage region.
\citet{Centeno2007} and \citet{Ishikawa2008} show examples of the emergence of THMFs both in the quiet Sun and in a plage region. In addition, \citet{Tsuneta2008ApJ} shows the ubiquitous occurrence of THMFs in polar regions.
These horizontal magnetic fields are subject to the upflows and downflows of convection.
Earlier reports of flux emergence in the quiet Sun \citep{Lites1996,DePontieu2002,MartinezGonzalez2007}
appear to be examples of THMFs.

Plage regions and the quiet Sun differ considerably in their amounts of vertical magnetic flux.
The properties of THMFs in plage regions and in the quiet Sun, however, appear to be similar \citep{Centeno2007,Ishikawa2008}.
A fundamental question is whether there is any difference in the properties of THMFs in these regions.
In this paper, we compare THMFs in a plage region with those in a quiet Sun region
to analyze their similarities and differences, and to clarify any relationship between the granular-sized THMFs and the existing vertical magnetic field. Relationships between THMFs and the larger-scale predominately vertical fields associated with plage and network regions may shed light on whether local or global-scale dynamo processes generate these transient fields. 
At present, the evidence that a local dynamo is playing a significant role for the quiet Sun magnetism
comes from the Hanle-effect investigation by \citet{Trujillo2004}, who inferred a magnetic energy density which is the order of 20\% of the kinetic energy density produced by the convective motions in the quiet solar photosphere, and
showed that the observed scattering polarization signals do not seem to be modulated by the solar cycle. 
In this paper we try to provide more evidence in favor of a local dynamo by comparing \emph{Hinode} observations of THMFs in a plage region and in the quiet Sun.

\begin{figure}
\includegraphics[height=8cm]{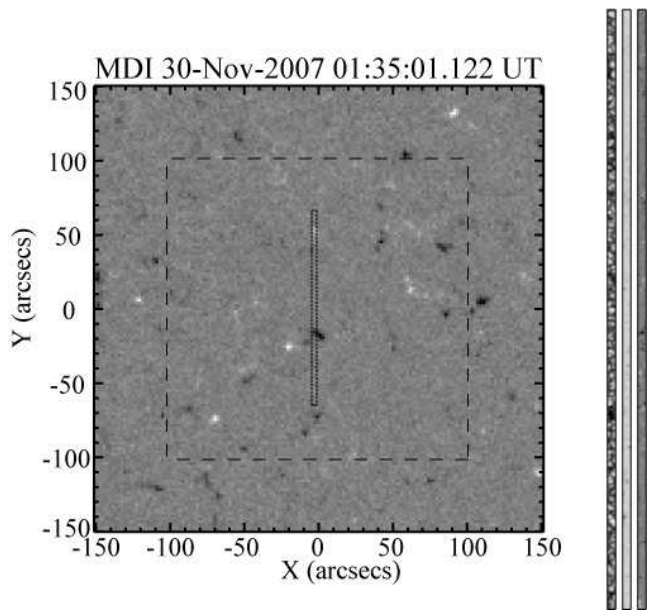}
\caption{$SOHO$ MDI longitudinal magnetograph taken at 01:35:01~UT on 2007
November~30. The approximate extent of the area scanned by the SP is shown by the rectangular box.
We did not get simultaneous magnetgorams with \emph{Hinode}, and cannot tell the exact position of scan.
The larger square shown with dashed lines shows the area used for estimate of total vertical flux.
The three images to the right represent the continuum intensity, the net linear polarization $LP$, and circular polarization of the slot obtained almost simultaneously with the magnetograph.}
\label{QSmdi}
\end{figure}

\begin{figure}
\includegraphics[height=8cm]{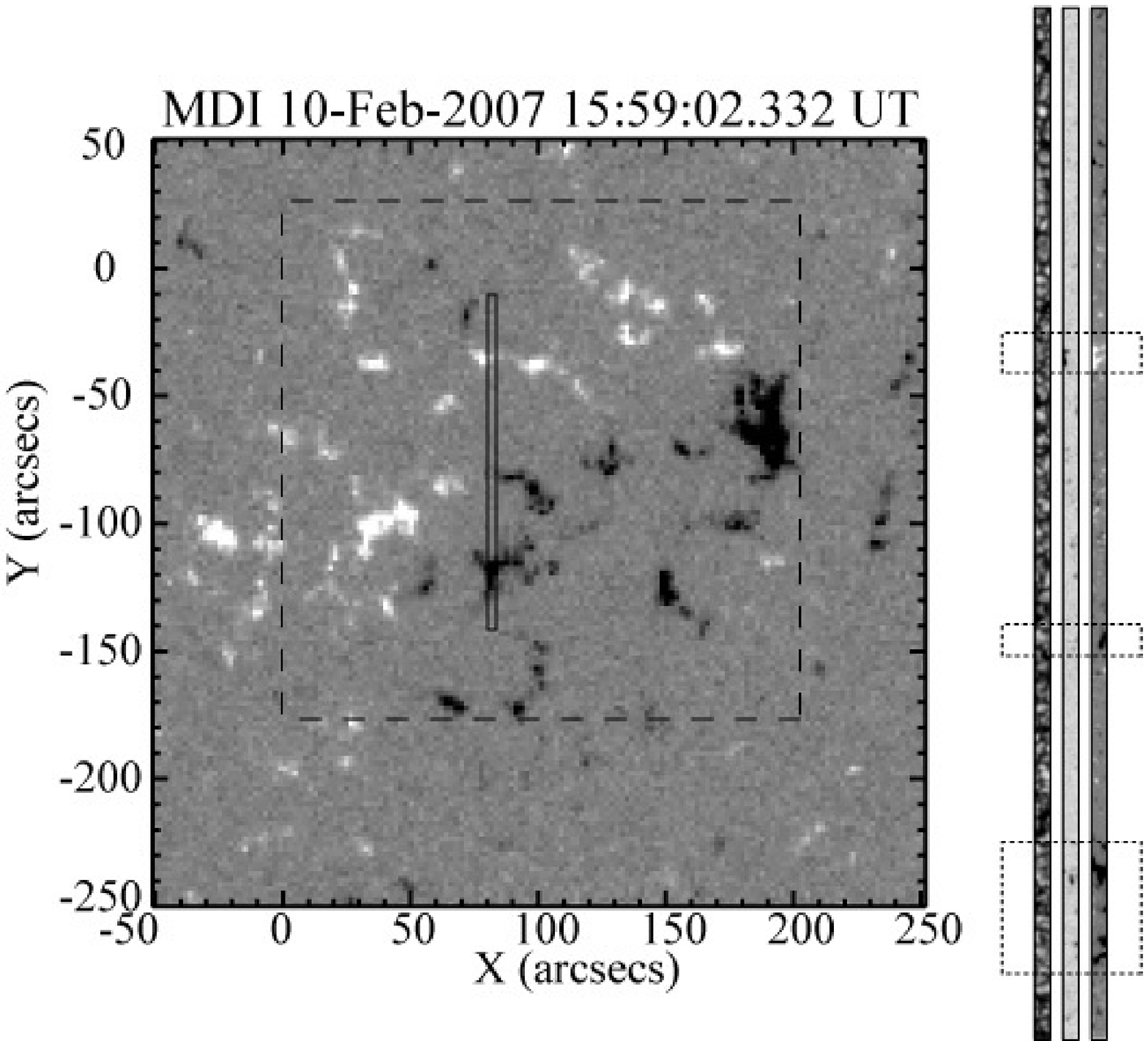}
\caption{$SOHO$ MDI longitudinal magnetograph taken at 15:59:02~UT on 2007
February~10. 
The caption is the same as Fig.\ref{QSmdi}.
Note that three rectangles with dotted line in slot images show the area which is dominated by persistent strong vertical magnetic flux concentrations for the plage region, and is masked for the analysis.
}
\label{PRmdi}
\end{figure}

\section{Observations}
A quiet region (Fig.\ref{QSmdi}) of $2\farcs0\times164\arcsec$ and a plage region (Fig.\ref{PRmdi}) of $2\farcs4\times164\arcsec$ were observed
with cadences of 28 and 34 seconds, respectively using the SP in Fast-Map mode.
The plage region was located at
($W72\arcsec, S76\arcsec$) heliographic coordinates, and was observed from 16:00 - 16:40 UT and 17:10 - 17:56 UT on 2007 February 10.
Subsequently a quiet solar region at disk center was observed on 2007 November 30 from 01:10 - 02:35 UT.

The SP produces Stokes spectral profiles ($I$, $Q$, $U$ and $V$) of the Zeeman-sensitive photospheric
Fe~I 630.15~nm and Fe~I 630.25~nm lines. The SP wavelength sampling is 2.15~pm pixel$^{-1}$.
The profiles at two successive slit positions, each with an integration time
 of 1.6~s, are summed. The spectra are also summed by 2~pixels along the slit.
This results in high cadences for the rectangular regions shown in Figs.~\ref{QSmdi} and \ref{PRmdi} with a pixel size of 0\farcs30 for the EW and 0\farcs32 for the NS direction.
Note that the noise level of Stokes $Q$ and $U$ on a single wavelength sample is $9.0\times10^{-4}I_{c}$, and that of Stokes $V$ is $8.0\times10^{-4}I_{c}$.

We use maps of the net linear polarization, $LP=\int \sqrt{Q^{2}+U^{2}}d\lambda/\int I_{c} d\lambda$.
The integration is carried out from -21.6~pm to 21.6~pm from the mean
center of the Stokes~$I$ profiles for Fe~I 630.25 nm without polarization signals.
The $LP$ maps both in the plage region and the quiet region show ubiquitous
THMFs are appearing and disappearing with a typical lifetime of 1-10~min, which is on the order of the lifetime of granules \citep{Ishikawa2008}.

\section{Comparisons between plage region and quiet Sun THMFs}
\subsection{Occurrence rates and position angle distribution}
\begin{figure}
\includegraphics[width=9cm]{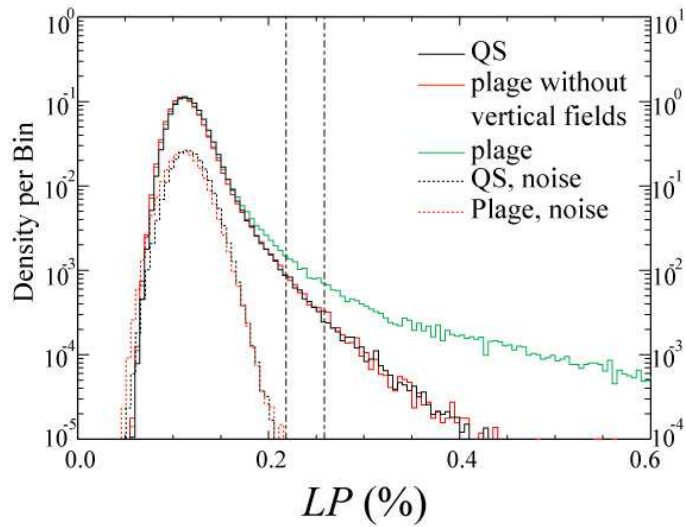}
\caption{Histograms of net linear polarization ($LP$) for plage and quiet Sun.
The dotted lines represent $LP$ noise distributions for both datasets. 
The noise is derived from the same formula as $LP$ but the integration is carried out at wavelengths far away from line center, without any Stokes signals.
The area dominated by vertical magnetic fields is masked in the plage region (see Fig.\ref{PRmdi}).
The two vertical dashed-dotted lines indicate $LP$ of 0.22\% and 0.26\% -- the thresholds used to identify THMFs in the datasets.}
\label{LPhist}
\end{figure}

\begin{figure}
\begin{center}
\includegraphics[width=6.5cm,angle=90]{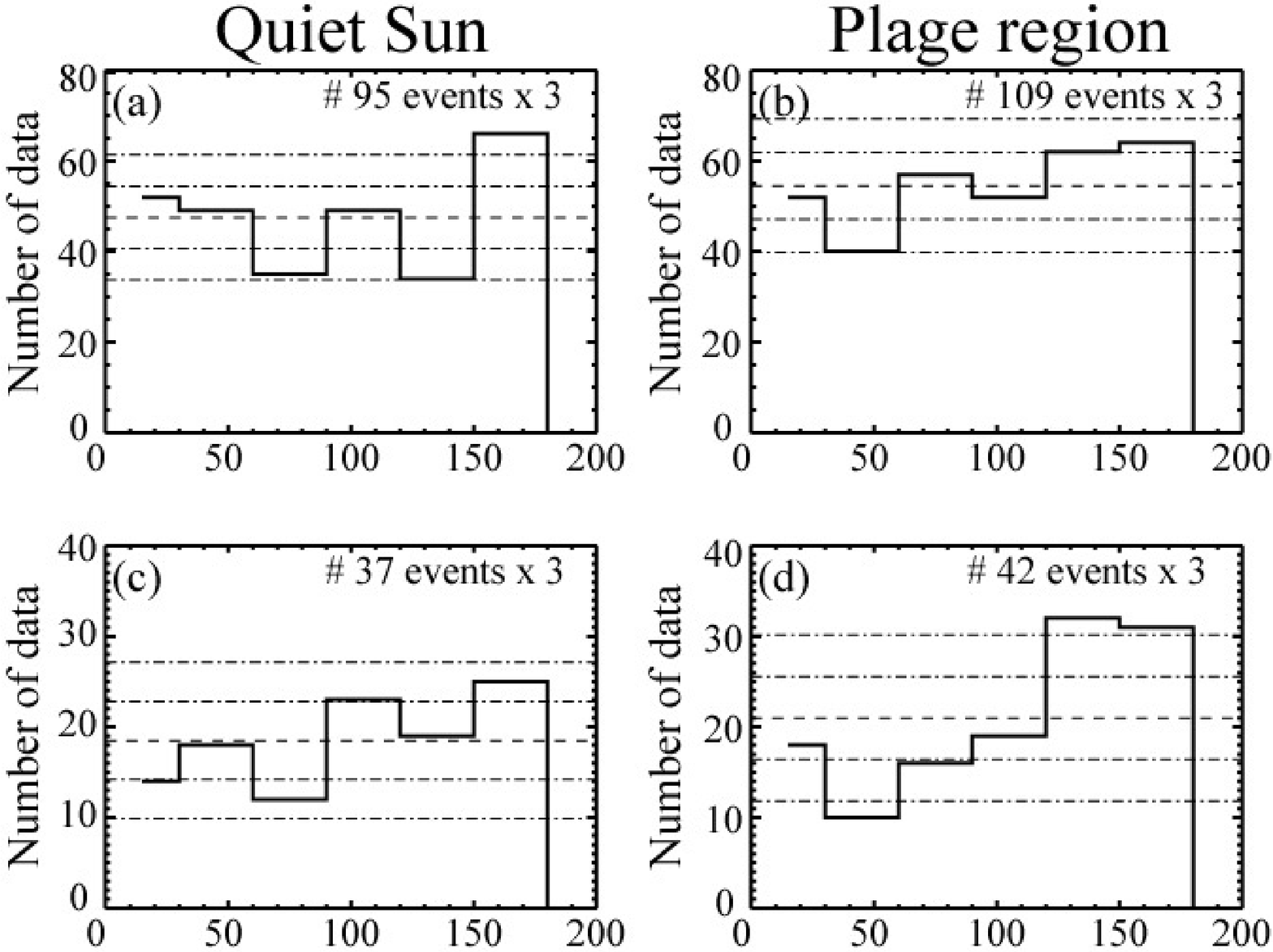}
\end{center}
\caption{
The histograms of the field azimuth angles for THMFs:
Three pixels with highest $LP$ are taken from individual events, and number distributions of
the azimuth angles of horizontal magnetic field for these pixels are plotted.
The azimuth angle 0$^\circ$ is to the West (right), 90$^\circ$ to
the North(up), and 180$^\circ$ to the East (left) (see Figs.\ref{QSmdi} and \ref{PRmdi}).
$(a)$ and $(b)$:Histograms of azimuth angle for 95 events in the quiet Sun and 109
events in the plage region, which have $LP$ higher than 0.22\%.
$(c)$ and $(d)$:Histograms of azimuth angle for 37 events in the quiet Sun and 42
events in the plage region, which have $LP$ higher than 0.26\%.
The dashed lines indicate the case for a uniform distribution.
Dashed-dotted lines closer to the dashed line show $\pm \sigma$, statistical deviation, and two other dotted lines show $\pm$ 2$\sigma$.
}
\label{azimuthhist}
\end{figure}

Figure \ref{LPhist} shows histograms of $LP$ for the plage as well as the quiet Sun region.
For the plage region 18\% of the observing area is occupied with persistent vertical magnetic flux concentrations for a period longer than the observing time of 86 min; we remove these area (Fig. \ref{PRmdi}) and plot the distribution with the red line (plage without vertical fields).
In the quiet Sun we do not see similarly stable vertical magnetic flux concentrations that persist during the entire observation time.
The distributions peak at $LP \sim$0.1\% and
the noise distributions show that the lower end of the distributions is dominated by the photon noise. 
Although the plage region, including the persistent vertical field region has much higher density in the higher end, 
the plage without persistent vertical fields and the quiet region have remarkably similar $LP$ distributions.  

We study only those THMF events that have $LP > 0.22\%$ and area $A \ge 3$~pixels.
We detect 95 events in 85~min in the quiet Sun and 109 events in 86~min in the plage region; we reject events found in the persistent strong magnetic concentrations in the plage region (Fig. \ref{PRmdi}).
The occurrence rate in the plage region is $1.2\times10^{-4}$ Mm$^{-2}$ s$^{-1}$, and thus this is nearly the same as  that in the quiet Sun region ($1.1\times10^{-4}$ Mm$^{-2}$ s$^{-1}$).
The remarkably similar $LP$ distributions in Fig. \ref{LPhist} also suggest the same occurrence rates of both the quiet Sun and the masked plage region. 
These occurrence rates are extremely high as discussed by \citet{Ishikawa2008}.
\footnote{The movie which shows the time evolution of transient horizontal fields is available at http://www.aanda.org. In the movie the yellow shows region with net linear polarization (LP) higher than 0.22\% ($\sim$155G in the weak field approximation) (we identify these regions as transient horizontal magnetic fields), and the green contours show the region with circular polarization larger than 0.3\% ($\sim$13G in the weak field approximation). We can see a lot of horizontal fields appear and disappear with granules. }

Figures \ref{azimuthhist} $(a)$ and $(b)$ show the magnetic field azimuth of THMFs 
for the 95 events in the quiet Sun and the 109 events in the plage region, which meet the threshold of $LP > 0.22\%$.
We use the azimuth angle from the ratio of wavelength-integrated $Q$ and $U$ ($\chi = \frac{1}{2}\arctan\left(\int Ud\lambda/\int Qd\lambda\right)$) \citep[see][chapter 11]{Landi2004} instead of
the Milne-Eddington inversion. 
The integration is carried out between -21.6 pm and -5.32 pm and between +5.32 pm and +21.6 pm from the center of Stokes $I$ profile for Fe~I 630.25~nm averaged over pixels without polarization signals.
The azimuth angle defined here has the usual 180$^\circ$ ambiguity.
We define $\sigma=\sqrt{N}$, where $N$ is a number of average events per 30$^\circ$ bin under the assumption of uniform azimuthal distributions of horizontal fields.
The azimuth angles except for the bin between 150$^\circ$ and 180$^\circ$ in the quiet Sun are essentially distributed within the 2$\sigma$ range.
It is difficult to conclude that the peak between 150$^\circ$ and 180$^\circ$ in the quiet Sun is significant, and there is no statistically significant orientation in either region.

Using a threshold of $LP > 0.26\%$ we detect 37 events in the quiet Sun 
and 42 events in the plage region. 
We show the azimuth angles of the magnetic vector for these THMFs in Figs.~\ref{azimuthhist} $(c)$ and $(d)$. 
In the plage region we find a broad peak between 120$^\circ$ and 180$^\circ$ and a dip between 30$^\circ$ and 60$^\circ$ that are significant at the 2$\sigma$ level. In contrast, these events in the quiet Sun still show an azimuth angle distributed within 2$\sigma$ of the uniform value. This peak angle corresponds to the tilt angle of the bipolar plage region line (Fig.\ref{PRmdi}). 
This indicates that THMFs with higher $LP$ in the plage region appaer to be partially related to the plage region's global fields.

\begin{figure}
\includegraphics[width=3.4cm,angle=90]{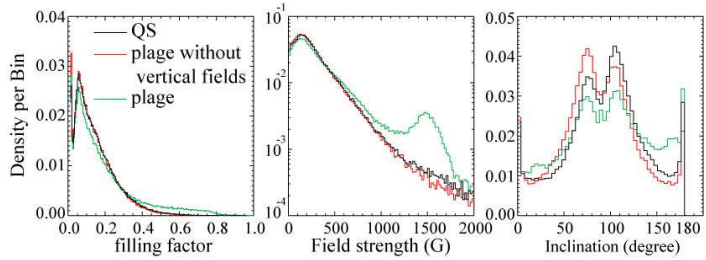}
\caption{
Filling factor ($left$), magnetic field strength ($middle$), and inclination ($right$) probability density functions (PDFs).
The black, green, and red lines represent the quiet Sun, the plage region, and the plage region excluding the area with persistent vertical fields.
}
\label{PDF}
\end{figure}
 \subsection{Distribution of magnetic field strength and inclination}
In the previous section we showed that THMFs essentially have the same properties in the plage region and in the quiet Sun with the exception of high $LP$ THMFs.
However, we only study THMFs that have relatively large $LP$. 
The threshold potentially excludes THMFs with smaller $LP$.
In order to investigate this issue further we derive probability density functions (PDFs) of intrinsic magnetic field strength and inclinations and compare PDFs between in the plage region and the quiet Sun. 
 
We employ a least-squares inversion based on a Milne-Eddington atmosphere to obtain the magnetic field vector and other parameters such as line-of-sight velocity, Doppler width, and magnetic filling factor $f$ from the Stokes spectra. We obtain a common stray light intensity profile from the Stokes $I$ profile averaged over pixels with a total degree of polarization less than 0.2\%.
We analyze pixels with the amplitudes of Stokes $Q$, $U$, or $V$ larger than 4.5 times their respective noise levels.
The pixels with $f < 0.01$ are excluded because Stokes profiles with such a small $f$ tend not to be fitted well.
  
\begin{figure}
\includegraphics[width=5cm,angle=90]{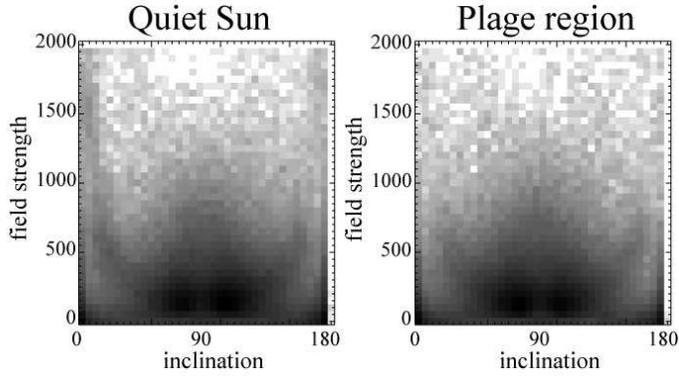}
\caption{
Scatter plots of inclination versus field strength for the quiet Sun ($left$) and the plage region without persistent vertical fields ($right$).
The gray scale indicates the logarithmic number of data points.
Most magnetic fields weaker than $\sim 500$ Gauss are horizontal for both regions.
}
\label{IF}
\end{figure}

\begin{figure}
\begin{center}
\includegraphics[width=5cm,angle=90]{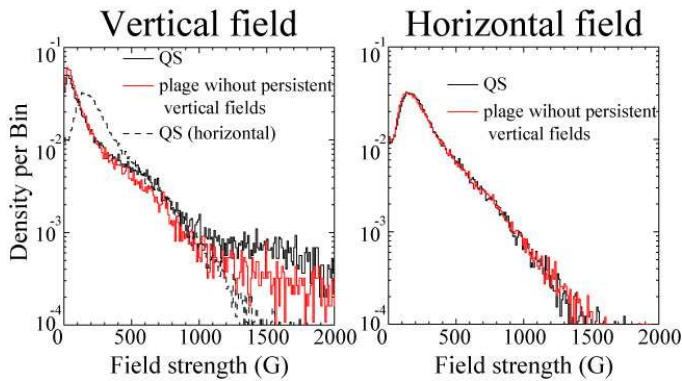}
\end{center}
\caption{
Magnetic field strength PDFs of vertical fields ($left$) and horizontal fields ($right$).
Black and red lines stand for the quiet Sun and the plage region without persistent vertical fields, respectively.
Black dashed line in $left$ figure shows PDF for plage-region's horizontal fields,
and is same as black solid line in $right$ figure.
Vertical fields refer to magnetic fields with inclination smaller than 20$^\circ$ or larger than 160$^\circ$,
and horizontal fields refer to magnetic fields with inclination larger than 70$^\circ$ and smaller than 110$^\circ$.
}
\label{VH_PDF}
\end{figure}

Figure~\ref{PDF} shows PDFs for filling factor, magnetic field strength and inclination.
The PDFs for the quiet Sun represented by the black line are considering the $\sim180000$ pixels.
The peaks of all PDFs for $f$ are located around 0.06. 
The analysis by \citet{Orozco2007} finds that quiet Sun magnetic fields have small filling factors with a peak at $f = 0.2$.
The discrepancy between our result and the value cited above may be due to our double pixel size ($\sim 0\farcs3$).
However, the important point here is that 
when we avoid the persistent vertical flux concentrations shown in Fig. \ref{PRmdi} for the plage region, the PDFs of $f$ for the quiet Sun and the plage region are remarkably similar.

The field strength ($B$) distribution with a peak of 100--200~Gauss of the plage excluding persistent vertical fields is again similar to that of the quiet Sun.
On the other hand, the plage region without mask also has a peak at around 150~Gauss, but its PDF decreases more slowly, and has a pronounced hump at around 1500~Gauss.
\citet{Orozco2007} have derived a similar PDF of field strength with two peaks for the quiet Sun, which in this case consists of both IN and network regions.
%\textbf{
Our FOV in the quiet Sun does not contain network fields, and corresponds to the inter network quiet Sun. 
Therefore, the PDF of field strength with the black line in the middle of Fig.~\ref{PDF} in this paper should be compared with that of the IN regions of Fig. 3 in \citet{Orozco2007}.
These two PDFs appear to be quite similar. 
Our quiet Sun PDF of field strength
%} 
has the slightly higher density between 1300 and 1800~Gauss compared with the plage region without persistent vertical flux concentrations.
This is due to a contamination of fragments of vertical magnetic fields similar to network fields in the quiet Sun during the observation.
It is clear that the hump around 1500~Gauss in the field strength PDF (Fig. \ref{PDF}) represented by the green line are due to plage-region's strong vertical network fields.
The PDFs for inclination ($\theta$) (Fig~\ref{PDF}, $right$) show that most of magnetic fields seem to be nearly horizontal in both plage and quiet regions.
Figure~\ref{IF} also indicates that weak magnetic fields ($B<\sim500$ Gauss) are nearly horizontal ($60^\circ < \theta < 120^\circ$) in both regions 
(a few weak magnetic fields are vertical $\theta \sim 0^\circ, 180^\circ$).

Next we derive field strength PDFs for vertical fields ($\theta < 20^\circ$ or $\theta > 160^\circ$) and horizontal fields ($70^\circ < \theta < 110^\circ$).
Vertical field distributions (Fig.~\ref{VH_PDF}, $left$) in both  quiet Sun and plage regions
extend to higher magnetic fields ($B>\sim1000$ Gauss) compared with horizontal magnetic fields represented by the black dashed line in Fig.~\ref{VH_PDF}, $left$.
We see that quiet Sun has a somewhat higher distribution towards above 500 G than the plage region.
This is likely caused by network fields in the quiet Sun, which are not removed. 
The same effect is seen in the PDF of field strength for the quiet Sun and the plage region without vertical fields (Fig.~\ref{PDF} ($middle$)).

The peaks of the PDFs (Fig.~\ref{VH_PDF}, $right$) for the horizontal field strength for both regions are located at about 150~Gauss, 
i.e., at the same location as the total field strength PDFs of Fig.~\ref{PDF}, $middle$.
Unlike vertical fields, the PDFs for horizontal fields in the quiet Sun and in the plage region are exactly the same.
The results suggest that horizontal fields characterized by THMFs with weak field strength dominate solar magnetic fields except for plage's global vertical magnetic field regions in terms of number density, and that they are uniform and universal in the solar surface.
Moreover, 93\% of horizontal magnetic fields have field strengths smaller than 700~G, and 98\% smaller than 1 kG for both regions.
Magnetic field strength of 700~Gauss corresponds to the typical equipartition field strength at the level of granules, 
where the density is $\sim10^{-6}$ g cm$^{-3}$ at the depth of 500 km and the velocity is 2 km s$^{-2}$.
Thus, majority of horizontal fields have field strengths smaller than the equipartition field strength for average granular flows.

We note that the peaks in the PDFs in Figs. \ref{PDF} and \ref{VH_PDF} may be artifacts; There still may be more abundant weaker fields, which we cannot see with our current threshold and sensitivity.
The small filling factors \citep[see also][]{Orozco2007} suggest that what \emph{Hinode} is now seeing may still be unresolved and be only the tip of the iceberg of the Sun's hidden magnetism inferred by \citet{Trujillo2004} through theoretical analysis of Hanle-effect observations.
We stress that THMFs with relatively weak fields ($B<$ 700 G) dominate the solar surface, and that the field strength distribution is exactly the same between the plage and quiet regions.

\section{Discussions and conclusions}
\subsection{THMF and local dynamo process}
The vertical magnetic flux in the plage is about 8 times larger than that of the quiet Sun.
Note that we derive these magnetic flux estimates from pixels with field strength stronger than $50$~Gauss in the regions marked by the boxes with dashed lines in Figs.~\ref{QSmdi} and \ref{PRmdi}.
Horizontal fields dominate the plage region without persistent large vertical magnetic flux concentrations as well as the quiet Sun, and
there is no difference in magnetic field strength of horizontal fields between these regions.
In addition the occurrence rates of THMFs are nearly identical.
If the THMF occurrence rate was in any way directly related to the \textit{global} vertical fields forming the plage region,
then we would expect the occurrence rate in the plage region to be much larger than that of the quiet Sun. 
This suggests that the emergence of the THMFs does not have a direct causal relationship with the vertical magnetic fields in the plage region.
The same THMF occurrence rate, no preferred orientation, and similar field strength distributions for both regions strongly suggest that a common local process that is not directly influenced by global magnetic fields produces THMFs.
As ubiquitous THMFs are receptive to convective motion \citep{Centeno2007,Ishikawa2008},
a reservoir for THMFs may be located near solar surface,
and these magnetic fields are carried to the surface with convective flow.

Such reservoir can be maintained by a local dynamo process due to near-surface convective motion \citep{Cattaneo1999,Vogler2007}. 
Indeed, numerical simulations \citep{Abbett2007,Schussler2008} have shown that local dynamo action can generate horizontal magnetic structures in the quiet Sun. 
Such a local dynamo process could naturally explain the similarity in occurrence rates and field strength PDFs, including the fact that THMFs do not have a preferred orientation.
Moreover, the horizontal field strength is distributed in a range smaller than the equipartition field strength of near-surface turbulent granules.
This would suggest that local dynamo process takes place in a relatively shallow layer with the size of a granule.
The similarity in field strength distribution also indicates that properties of THMFs do not depend on the seed field e.g. global fields.

Other possibilities for the origin of THMFs include debris from decaying active region, magnetic fields failed to emerge from the convection region to the photosphere \citep{magara2001}, and extended weak magnetic fields in the upper convection zone generated by \textit{explosion} \citep{moreno1995}.
If the reservoir is maintained by one of these processes the THMFs would be expected to be affected by global toroidal fields in terms of properties of THMFs described above.
\citet{Steiner2008} carried out numerical simulations of the surface layers of the Sun with two different sets of initial and boundary conditions, and compared their simulations with Hinode observations. 
The first set starts with vertical unipolar magnetic fields appropriate for plage regions, and the second set with horizontal uniform magnetic fields at the bottom boundary area.
In both runs they found that spatial and temporal average for the horizontal fields surpass that for the vertical fields but the first run has more vertical fields than the other. 
As these simulations show, the environment of magnetic fields seems to affect the distribution of the magnetic field vector.
Within the context of these simulations the  
observed properties of THMFs such as the similarity in the occurrence rates and magnetic field distributions, and no preferred orientation of THMFs would be difficult to explain.

A slight preferred orientation of THMFs with higher $LP$ toward the global plage polarity suggests that these THMFs may be influenced by the global plage field.
However, because any strong vertical fields associated with the emergence of these THMFs \citep{Ishikawa2008} are not observed, they are probably not directly created from the vertical magnetic fields forming the plage as suggested by \citet{Isobe2008}.
Thus, even if these THMFs with higher $LP$ are related to the global toroidal system,
the relationship would be indirect --- these THMFs with high $LP$ may result from fragmented elements of plage flux tossed about by the convective motions below the photosphere.

\subsection{THMF and chromospheric and coronal heating}
Vertical magnetic elements on the solar surface are naturally connected with the upper atmosphere.
It is essential for coronal and chromospheric heating problem to understand their properties and the connectivity with the chromosphere and the corona. 
This raises a parallel question for the case of the THMFs: do the ubiquitous THMFs in the quiet Sun's photosphere have any consequence on upper atmosphere? 
We here estimate the magnetic flux carried by THMFs as follows.
\begin{equation}
E_{mag}=\frac{B^{2}}{8\pi}\times V \times f \times \frac{1}{\tau} \times \frac{1}{A} 
\label{fluxeq}
\end{equation}
where $B$ is average magnetic field strength of THMFs ($\sim310$ G), $f$ average filling factor ($\sim 0.15$),  and $A$ the observing area  ($2\arcsec\times164\arcsec$) for the quiet Sun, 
and $B$ of $\sim370$ G, $f$ of $\sim 0.18$, and $A$ of $2\farcs4\times164\arcsec\times0.84$ for the plage region.
In order to obtain field strength and filling factor of THMFs, the Milne
Eddington inversion was performed for pixels with Stokes $Q$, $U$, or $V$
amplitudes 4.5 times larger than their noise levels (section 3.2). Since the
selected pixels would include vertical flux tubes, we further set the
condition that $LP$ be higher than 0.22\%, which is
threshold used for detecting THMFs (section 3.1). Average field strength
and filling factor of the pixels thus chosen are used for $B$ and $f$ in
equation (\ref{fluxeq}).
THMFs are at most as large as a granule where they appear \citep{Ishikawa2008}, and the size is comparable to that of a granule \citep[][in prep.]{Ishikawa2008b}.
We substitute $\sim(1\arcsec)^{3}$ for the
volume ($V$).
THMF appears in the quiet Sun and in the plage region every 54 s and 47 s as described in previous section, and we substitute 54 s and 47 s for $\tau$, respectively.
Then, we obtain $\sim2\times10^{6}$ erg cm$^{-2}$ s$^{-1}$ and $\sim5\times10^{6}$ erg cm$^{-2}$ s$^{-1}$ for the quiet Sun and the plage region, respectively.
Note that the occurrence rate may be under-estimated due to our limited sensitivity. Thus, the estimated energy flux should be taken as a minimum value.

\begin{table}
\begin{center}
\begin{tabular}{lccc}
\hline
\hline
\multicolumn{1}{c}{Energy loss} (erg cm$^{-2}$ sec$^{-1}$) &Quiet Sun&Active region\\ \hline
\ \ Corona &\mbox{$3\times10^{5}$}&\mbox{$10^{7}$}\\
Chromosphere &\mbox{$4\times10^{6}$}&\mbox{$2\times10^{7}$}\\ \hline
Magnetic energy &Quiet Sun&Plage region&\\
supplied by THMFs  &\mbox{$\sim2\times10^{6}$}&\mbox{$\sim5\times10^{6}$}\\
\hline
\end{tabular}
\end{center}
\caption{Chromospheric and coronal energy losses \citep{withbroe&noyes1977} and magnetic energy flux (bottom) in the quiet Sun and in a plage deduced in this paper}
\label{EL}
\end{table}

Table \ref{EL} shows energy losses of active regions and quiet regions obtained from the empirical models \citep[][Table 1]{withbroe&noyes1977}. Derived estimates of energy losses in each region indicate the required energy input. The magnetic energy flux that THMFs provide to the quiet Sun ($\sim2\times10^{6}$ erg cm$^{-2}$ s$^{-1}$) is comparable to total chromospheric energy loss of 4$\times10^{6}$ erg cm$^{-2}$ s$^{-1}$. The total energy loss of the corona is only one tenth of the energy loss in the chromosphere. The magnetic energy flux in a plage region of $\sim5\times10^{6}$ erg cm$^{-2}$ s$^{-1}$ associated with THMFs is comparable to the total chromospheric and coronal energy loss. 

We demonstrated that the horizontal magnetic fields could contribute to the chromospheric and coronal heating \citep[e.g.,][]{Abbett2007,Isobe2008}. Its transient nature may accompany an associated transient chromosheric heating and dynamic events such as many small-scale dynamical events as shown by the \textit{Hinode} CaII H movies.  We point out a possible connection between the horizontal magnetic fields and high chromospheric activities such as jets discovered by \textit{Hinode}/SOT \citep{shibata2007}. We also indicate that the THMFs  we observe are probably the tip of the ice berg due to our limited sensitivity, and there may be more ubiquitous magnetic fields unresolved by \textit{Hinode} \citep{Trujillo2004}.

\begin{acknowledgements}
The authors thank T. E. Berger, A. Ferriz Mas, H. Isobe, T. Rimmele, and T. D. Tarbell for stimulating discussion and editorial help.
The authors also thank the Japan-U.S.A. SOT team for their outstanding work in the construction and operation of SOT.
\emph{Hinode} is a Japanese mission developed and launched by ISAS/JAXA, with NAOJ as domestic partner and NASA and STFC (UK) as international partners. It is operated by these agencies in co-operation with ESA and NSC (Norway). The National Center for Atmospheric Research is sponsored by the National Science Foundation.\end{acknowledgements}
\bibliographystyle{aa}
%\bibliography{ms}

\end{document}